\documentclass[reprint,amsmath,amssymb,aps,prb,twocolumn,superscriptaddress,showpacs,twoside,10pt,floatfix]{revtex4-1}
\usepackage[utf8x]{inputenc}
\usepackage{amssymb,amsmath}
\usepackage{graphicx}
\usepackage{color}
\usepackage[colorlinks,bookmarks=true,citecolor=blue,linkcolor=blue,urlcolor=blue,breaklinks=false]{hyperref}
\usepackage{bbm} 
\usepackage{subfigure}

\newcommand{\et}{{\it et al. }}
\newcommand{\bea}{\begin{eqnarray}}
\newcommand{\eea}{\end{eqnarray}}
\newcommand{\be}{\begin{equation}}
\newcommand{\ee}{\end{equation}}

\newcommand{\la}{\langle}
\newcommand{\ra}{\rangle}

\begin{document}
\title{Location of the Potts-critical end point in the frustrated Ising model\\on the square lattice}
\author{Ansgar Kalz}
\email{kalz@theorie.physik.uni-goettingen.de}
\affiliation{Institut für Theoretische Physik, Georg-August-Universität Göttingen, 37077 Göttingen, Germany}
\author{Andreas Honecker}
\affiliation{Institut für Theoretische Physik, Georg-August-Universität Göttingen, 37077 Göttingen, Germany}
\affiliation{Fakultät für Mathematik und Informatik, Georg-August-Universität Göttingen, 37073 Göttingen, Germany}
\date{September 12, 2012; published: October 11, 2012}

\begin{abstract}
We report on Monte Carlo simulations for the two-dimensional frustrated $J_1$-$J_2$ Ising model on the square lattice. Recent analysis has shown that for the phase transition from the paramagnetic state to the antiferromagnetic collinear state different phase-transition scenarios apply depending on the value of the frustration $J_2 / J_1$. In particular a region with critical Ashkin-Teller-like behavior, i.e., a second-order phase transition with varying critical exponents, and a noncritical region with first-order indications were verified. However, the exact transition point $[J_2/J_1]_C$ between both scenarios was under debate. In this paper we present Monte Carlo data which strengthens the conclusion of Jin \et [Phys. Rev. Lett. \textbf{108}, 045702 (2012)] that the transition point is at a value of $J_2/J_1 \approx 0.67$ and that double-peak structures in the energy histograms for larger values of $J_2/J_1$ are unstable in a scaling analysis.
\end{abstract}


\maketitle

\section{Introduction}
The introduction of competing interactions in the classical two-dimensional $J_1$-$J_2$ Ising model is accompanied by the appearance of new ground states and critical points at which the ground state shows a large degeneracy. Additionally ,the frustration can also affect the critical behavior of the model. In particular, the emergence of varying critical exponents for the transition from a high-temperature paramagnetic phase to a collinear phase was observed numerically.\cite{P:lanbin80, B:lanbin00}

However, a scenario of a continuous phase transition with varying exponents was in question after mean-field calculations by Morán-Lopéz \et\cite{P:lopez93, P:lopez94} gave evidence for a first-order transition in a certain regime of the frustration. The first-order scenario was strengthened by Monte Carlo simulations which mainly focused on the evaluation of energy histograms in the vicinity of the transition temperature.\cite{P:kalz08, P:kalz09} 

Recently, we presented in Ref.\,\onlinecite{P:kalz11b} a full analysis for the phase transition using conformal field theory and extensive Monte Carlo simulations. The focus was on the derivation of the underlying field theory starting at the point of two decoupled Ising models at $J_1=0$, which is represented by a $c=1$ field theory. By including the nearest-neighbor coupling $J_1$ perturbatively, we arrived at an Ashkin-Teller field theory which has a central charge of $c=1$ and is known to exhibit varying critical exponents.\cite{B:ginsparg88} Moreover, the Potts-critical end point of this theory allows for the onset of a noncritical phase transition\cite{P:ditzian80, B:baxter82} and could thereby explain the two different scenarios observed in the frustrated Ising model. The energy histograms showed two-peaked structures in the intermediate regime $J_1/2 < J_2 \lesssim J_1$. This led to the conclusion that the position of the critical end point was at $J_1 \approx J_2$. Jin \et\cite{P:jin11} agree with the general picture of two different transition scenarios and an underlying Ashkin-Teller field theory but argue that the first-order behavior is only valid up to $J_2 \lesssim 0.67\,J_1$. Using mainly arguments about the universality of the Binder cumulants, they show the equivalence of the point $J_2 \approx 0.67\,J_1$ in the frustrated Ising model and the four-state Potts model,\cite{P:wu82} which marks the critical end point of the Ashkin-Teller field theory.\cite{B:ginsparg88, P:ditzian80, B:baxter82} This universality for absolute values of the Binder cumulants was discussed extensively, e.g., for anisotropic couplings and different lattice systems in the Ising model and holds only under certain conditions.\cite{P:chen04, *P:chen05, P:selke05, *P:selke06}

On the other hand, while the finite-size dependence of the histogram shape was analyzed for small parameters $J_2 = 0.6\,J_1$ and $0.65\,J_1$,\cite{P:kalz08, P:kalz09} such an analysis was not performed for $J_2 \geq 0.7\,J_1$ due to the increasing length scales needed to exhibit the double-peak structure in the first place. Thus, the onset of doubly peaked features in the histograms was interpreted as a signal for a first-order transition. 
In this work we present a finite-size analysis for the energy histograms on a larger scale and observe the vanishing of first-order signals for $J_2 \gtrsim 0.67 \,J_1$, in agreement with the conclusion of Jin \et\cite{P:jin11} that the first-order scenario ends at $J_2 \approx 0.67 \, J_1$. Furthermore we analyze the trend of the critical exponents in the intermediate regime $0.67\,J_1 < J_2 \leq 1.2\,J_1$ to verify the convergence of the Ashkin-Teller model and its exponents to the Potts-critical end point.

After a short introduction to the model in Sec.\,\ref{s:model} we present Monte Carlo results for histograms and critical exponents in Sec.\,\ref{s:results}; in particular the scaling analysis will be presented as an example at the point $J_2 = 0.8\,J_1$. We summarize our results in the concluding Sec.\,\ref{s:dis}.

\section{Model}\label{s:model}
The frustrated Ising model is described by the classical Hamiltonian
\bea
H_{\text{Ising}} = J_1 \sum_{\text{NN}} S_i S_j + J_2 \sum_{\text{NNN}} S_i S_j 
\eea
which sums over all antiferromagnetic ($J_1,\,J_2 > 0$) nearest-neighbor (NN) and next-nearest-neighbor (NNN) interactions of spin variables $S_i=\pm 1$ on a square lattice.

\begin{figure}[t!] 
\includegraphics[width=0.48\textwidth]{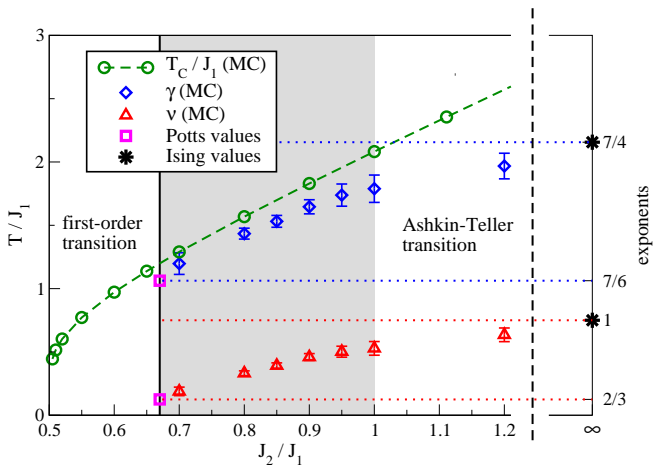}
\caption{\label{f:phase} (Color online) Phase diagram for the frustrated Ising model for $J_2 > J_1/2$. Two different regions are identified where the phase transitions show different behavior. In particular the shaded area in the middle is discussed in this work by means of histograms and critical exponents ($\nu$, red triangles; $\gamma$, blue diamonds). Data for the critical temperatures (green circles) are from Ref.\,\onlinecite{P:kalz11b}.}
\end{figure}

Ground-state configurations are given by a twofold degenerate Néel state for $J_2 < J_1/2$ and a fourfold degenerate collinear state for $J_2 > J_1 / 2$. At the critical point $J_2 = J_1/2$ the transition temperature is suppressed to zero, and a ground-state manifold with a degeneracy of linear order is present.\cite{P:kalz08} The phase transition to the Néel state belongs to the two-dimensional Ising universality class and will not be discussed any further. However, for $J_2 > J_1/2$ two different phase-transition scenarios apply for small and large values of frustration $J_2/J_1$: Ashkin-Teller like critical behavior with varying critical exponents for $J_2/J_1 \nearrow \infty$ and first-order noncritical phase transitions for $J_2/J_1 \searrow 0.5$.\cite{P:kalz11b, P:jin11} In Fig.\,\ref{f:phase} the critical temperatures are shown as green circles; the exact transition point between both regimes will be discussed in the following section, and the area in question is marked in gray.
\section{Monte Carlo Results}\label{s:results}
For the following numerical analysis we used a single-spin Metropolis update\cite{P:metropolis53}  with additional temperature exchange Monte Carlo steps\cite{P:hukushima96, P:hansmann97, P:katzgraber06} as it was also used in earlier works on the frustrated Ising model.\cite{P:kalz08, P:kalz09, P:kalz11b}
\begin{figure}[t!] 
\includegraphics[width=0.48\textwidth]{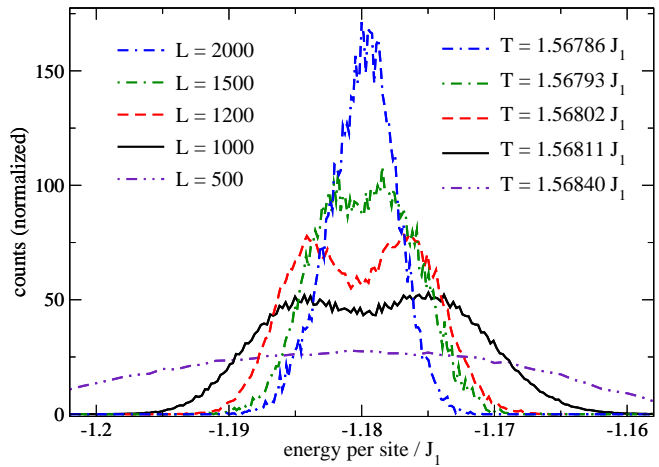}
\caption{\label{f:histo} (Color online) Histograms for different lattice sizes recorded for critical (size dependent) temperatures at $J_2 = 0.8\,J_1$. After the emergence of a double peak at intermediate system sizes $L = 1000,\,1200$ the vanishing of this first-order signature is observed for even larger systems ($L=1500,\,2000$).}
\end{figure}

To investigate the nature of the phase transition the first focus was on the computation of energy distributions. In Ref.\,\onlinecite{P:kalz11b} we presented a histogram at $J_2 = 0.8\,J_1$ for a lattice of linear size $L = 1000$ with a double-peak feature. This was interpreted as evidence for a first-order transition. For this work we computed systems with $L=500,\,1000,\,1200,\,1500,\,2000$ and present in Fig.\,\ref{f:histo} histograms at the size-dependent transition temperature. Surprisingly, the doubly peaked shape is observed only for intermediate system sizes $1000 \leq L \lesssim 1500$. These first-order signatures appear at a particular crossover scale, but the distance between the two peaks decreases and vanishes completely when the linear system size is doubled again ($L=2000$). A similar disappearance of a double-peak structure in energy histograms has been reported previously for the Baxter-Wu model.\cite{P:schreiber05} In that case, the two peaks approached each other in the thermodynamic limit but were present for all finite system sizes. In contrast, the present situation where a double-peak structure emerges first at intermediate system sizes $L \approx 1000$ and then disappears again is highly unusual. Due to the vanishing double-peak structure  the scenario of a first-order transition is no longer valid for the intermediate value of $J_2 = 0.8\,J_1$, and further analysis of the critical behavior is necessary.
\begin{figure}[t!] 
\subfigure[\label{sf:nu} extraction of $\nu$]{\includegraphics[width=0.48\textwidth]{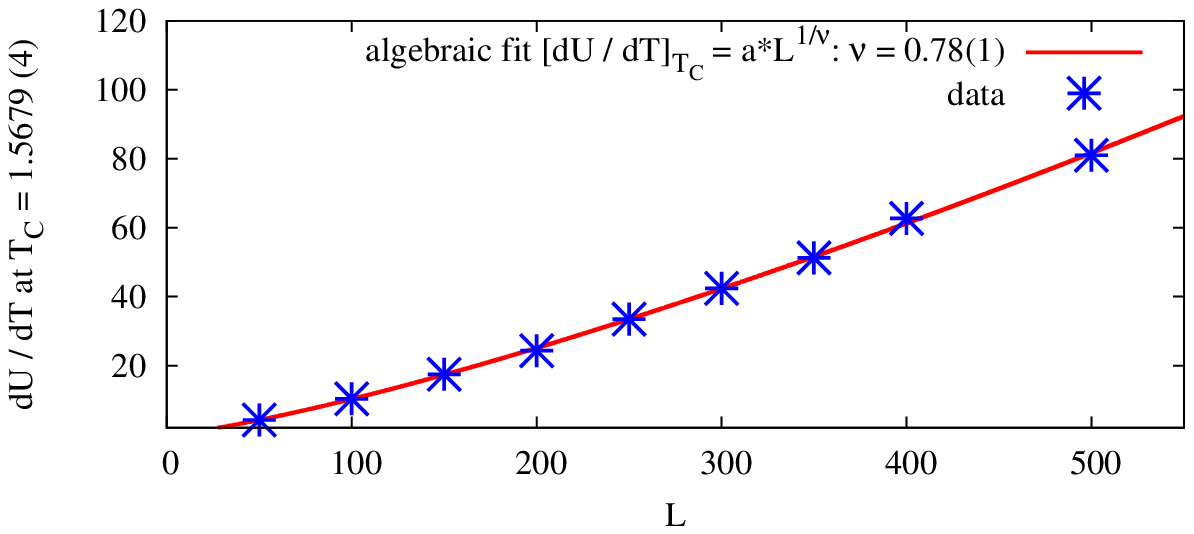}}\\
\subfigure[\label{sf:collapse} scaling collapse for $\nu$]{\includegraphics[width=0.48\textwidth]{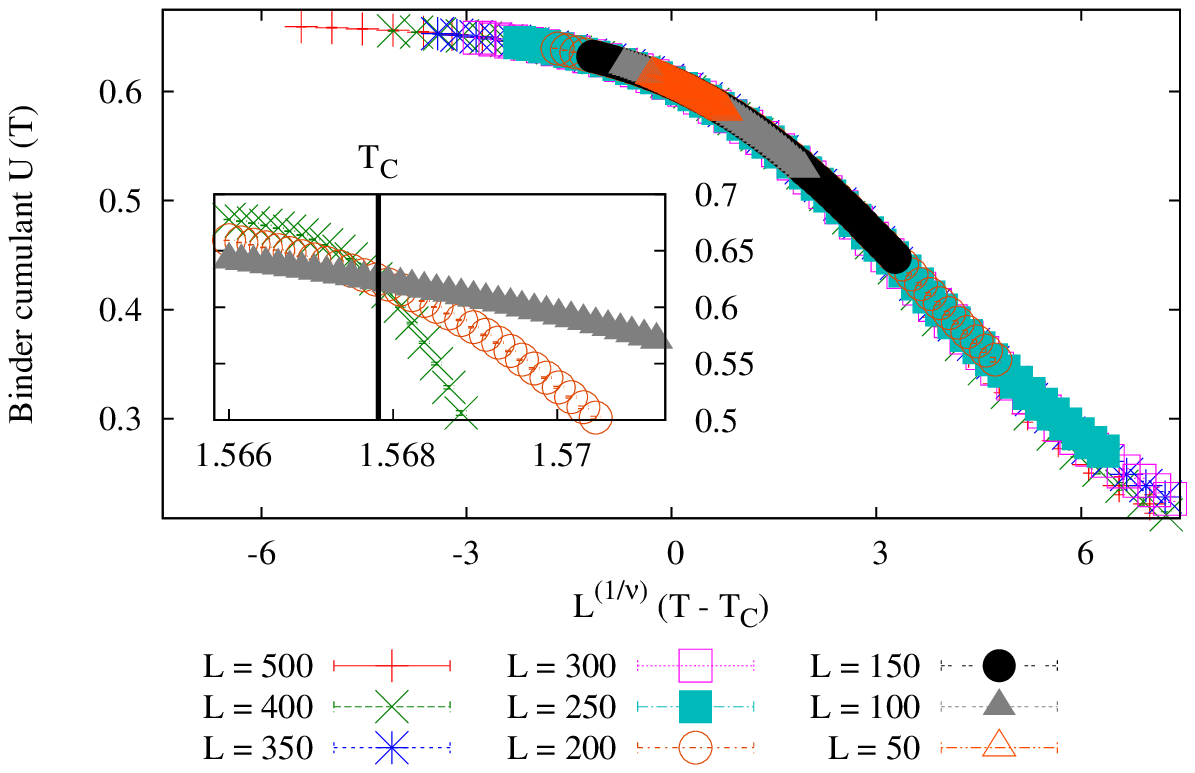}}\\
\subfigure[\label{sf:gamma} extraction of $\gamma$]{\includegraphics[width=0.48\textwidth]{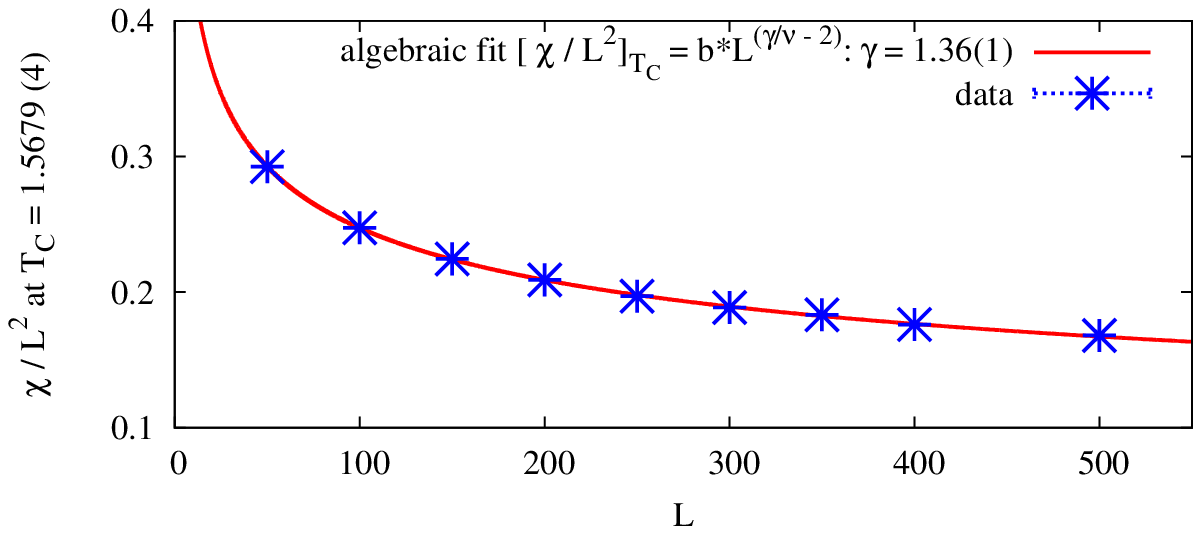}}
\caption{\label{f:expo} (Color online) Computation of critical exponents $\nu$ and $\gamma$ from the Binder cumulants and susceptibility of the magnetic order parameter at $J_2 = 0.8\,J_1$. (a) and (c) show size-dependent values of the derivative $[dU/dT]_{T_C}$ and susceptibility $[\chi / L^2]_{T_C}$ and algebraic fits that yield $\nu = 0.78(1)$ and $\gamma = 1.36(1)$. In (b), additionally, a scaling collapse of the Binder cumulant is shown for given $T_C$ and $\nu$; three original Binder cumulants are shown in the inset.}
\end{figure}

In light of the recent work by Jin \et \cite{P:jin11} we checked the development of critical exponents for varying frustration $J_2/J_1 \searrow 0.67$ by analyzing our Monte Carlo data for system sizes from $N = 50\times 50$ to $N = 500\times 500$. A detailed investigation of correlation functions and the corresponding exponent $\eta$ would not be meaningful since $\eta = 1/4$ is expected to be constant in the Ashkin-Teller model.\cite{B:baxter82} By applying the scaling relation $\gamma/ \nu = 2 - \eta$ the ratio of $\gamma / \nu = 7/4$ is also fixed.\cite{B:baxter82, P:jin11} However, the exponents $\gamma$ and $\nu$ can be extracted separately. In particular the Binder cumulants\cite{P:binder81L, P:binder81Z} $U_B$ and the susceptibility $\chi$,
\bea
U_B = 1 - \frac{\la m^4 \ra}{3\la m^2 \ra^2}\,, \quad \quad \quad \chi = \frac{\la m^2 \ra}{T\, L^2} 
\eea
for the collinear phase have been computed for this purpose. The order parameter is defined as
\bea
m = m_x + m_y\,, \quad m_{x,y} = \tfrac{1}{N} \sum_i (-1)^{i_{x,y}}S_i\,
\eea  
and satisfies $\la m \ra = 0$ for all temperatures in finite systems. The results were double-checked for some parameters with a slightly different definition of the order parameter $m^2 = m_x^2 + m_y^2$ which was used in Ref.\,\onlinecite{P:jin11}. In the scaling analysis for critical temperatures and exponents we did not notice any differences.
\begin{figure}[t!]  
\includegraphics[width=0.48\textwidth]{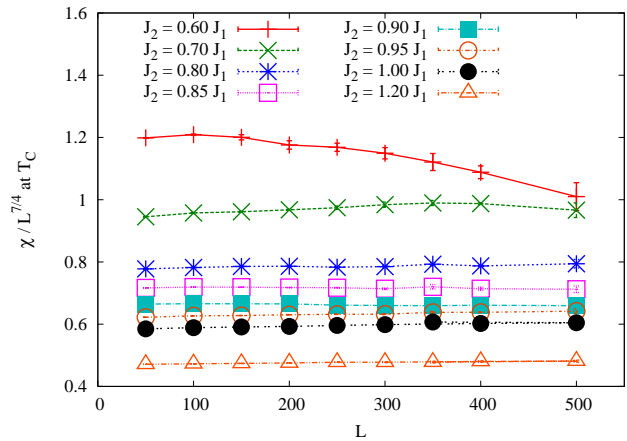}
\caption{\label{f:flow} (Color online) Flowgrams for the susceptibility $\chi$ at the transition temperature $T_C$ for several ratios $J_2/J_1$. Only for $J_2 \gtrsim 0.7\,J_1$ is the scaling nearly constant in $L^{-7/4}$ and Ashkin-Teller-like behavior obtained.}
\end{figure}

The critical exponent $\nu$ of the correlation length $\xi$ is calculated from the scaling of derivatives for Binder cumulants of different system sizes at the transition temperature:\cite{B:lanbin00, P:kalz08}
\bea 
dU_B/dT |_{T_C} = a\, L^{1/\nu}\,.
\eea 
Hereby we neglect scaling corrections which have only a small influence on the resulting exponent, i.e., the deviation is smaller than the fitting error. It should be noted that the analysis becomes more sensitive to corrections for $J_2/J_1 \rightarrow 0.67$. Nevertheless, the logarithmic corrections known to be present at the Potts point had no crucial influence in our analyses for $J_2/J_1 \geq 0.7$. As an example the scaling analysis is shown for $J_2 = 0.8\,J_1$ in Fig.\,\ref{sf:nu}: at the critical temperature $T_C = 1.5679(4)\,J_1$ a critical exponent $\nu = 0.78(1)$ is extracted. The same values are used in a finite-size-scaling collapse\cite{B:lanbin00, P:kuklov08, P:charrier08} of the Binder cumulants vs $L^{1/\nu}(T-T_C)$ in Fig.\,\ref{sf:collapse}, and the good agreement in a wide temperature and lattice-size regime verifies the assumption of criticality in general and confirms the extracted values in particular. The same procedure was performed for several ratios of $J_2/J_1>0.67$, and all values of $\nu$ are shown in Fig.\,\ref{f:phase} as red triangles. It is clearly observable that the exponents vary monotonically from the Ising value $\nu_{\text{Ising}} = 1$ for $J_1 = 0$ (right-hand side) towards the four-state Potts value\cite{P:wu82} $\nu_{\text{Potts}} = 2/3$ for $J_2/J_1 \searrow 0.67$ (straight black vertical line). A comparison with earlier works is ambiguous: Our values for $\nu$ and $J_2 \geq 0.7\,J_1$ are in good agreement with the data presented in Ref.~\onlinecite{P:lanbin85}, although our values have smaller errors. However, more recent analyses at the point $J_2=J_1$ yield a value of roughly $\nu = 0.84(1)$\cite{P:malakis06, P:yin09, P:badiev11} at $T_C =2.082(1)\,J_1$, whereas our analysis yields a value of $\nu = 0.88(2)$ for a critical temperature $T_C =2.0819(4)\,J_1$. This discrepancy shows the importance of the additional scaling-collapse analysis which helps to refine the critical temperature and exponent.

A similar scaling analysis is performed for the susceptibility and is also shown as an example at $J_2 = 0.8\,J_1$ in Fig.\,\ref{sf:gamma}. However, the critical exponent $\gamma$ also depends on the previously extracted $\nu$ and is obtained without scaling corrections via
\bea 
\chi |_{T_C} = b\, L^{\gamma / \nu}.
\eea
The development of $\gamma (J_2/J_1)$ is illustrated by the blue diamonds in Fig.\,\ref{f:phase}. The limits are given by $\gamma_{\text{Ising}} =7/4$, and $\gamma_{\text{Potts}} = 7/6$.\cite{P:wu82} Along the line of phase transitions the ratio of $\gamma$ and $\nu$ is roughly constant at $7/4$ which is valid for both the Ising model and the Ashkin-Teller model including the Potts-critical end point. This fact is also observable from the flowgram presented in Fig.\,\ref{f:flow}: We present the flowgram of the susceptibility for different parameters $J_2/J_1$ in addition to the scaling analysis. According to Refs.\,\onlinecite{P:kuklov06, P:kuklov08}, the behavior of the flow of some observable connected to the phase transition should change significantly if the type of this transition is altered. Such an alteration is visible in Fig.\,\ref{f:flow} for the flow at $J_2/J_1 = 0.6$, which is not constant in the chosen scaling over $L^{-7/4}$. For all other parameters, i.e., $J_2/J_1 \geq 0.7$ the flow of the susceptibility shows only a small deviation from a constant scaling, which fits into the picture of the Ashkin-Teller critical behavior.

\section{Discussion}\label{s:dis}
We presented Monte Carlo data on the paramagnetic-collinear phase transition in the frustrated Ising model. Our data are in  agreement with the recent findings of Jin \et \cite{P:jin11} and provide further support for the Ashkin-Teller nature of this phase transition in a specified parameter region. In particular we analyzed the evolution of the critical exponents $\nu$ and $\gamma$ from their Ising values (at $J_2/J_1 \nearrow \infty$) to the four-state-Potts values (at $J_2/J_1 \approx 0.67$). Thus, the critical frustration parameter of $[J_2/J_1]_C \approx 0.67$ proposed by Jin \et \cite{P:jin11} is consistent with our analysis. A comparison of the absolute values of $\nu$ with estimates given by Landau and Binder\cite{P:lanbin85} shows good agreement and smaller errors for our data. However, at the present level of accuracy the values deviate from recent findings of several other groups.\cite{P:malakis06, P:yin09, P:badiev11} We also presented a finite-size analysis for the energy distribution at $J_2 = 0.8\,J_1$ and showed that first-order signals in the histograms vanish for linear system sizes $L\gtrsim 1200$, and therefore the scenario of a critical phase transition is strengthened also by the absence of a latent heat.

\begin{acknowledgments}
We thank the Deutsche Forschungsgemeimschaft for financial support via the Collaborative Research Centre SFB 602 (TP A18). Furthermore, the Monte Carlo simulations were performed on the parallel clusters of the Gesellschaft für wissenschaftliche Datenverarbeitung Göttingen (GWDG) and the North German Supercomputing Alliance (HLRN), and we thank them for technical support. We acknowledge fruitful discussions with Anders Sandvik and Arnab Sen (authors of Ref.\,\onlinecite{P:jin11}) and Pierre Pujol from the University of Toulouse.
\end{acknowledgments}

\bibliography{../../IsingNote}

\end{document}